\documentclass[twocolumn,floatfix,prb,showpacs,preprintnumbers,amsmath,amssymb,superscriptaddress]{revtex4-1}

\usepackage[spanish]{babel}

\usepackage[latin1]{inputenc}
\usepackage{graphicx}
\usepackage{dcolumn}
\usepackage{bm}
\usepackage{xspace}
\usepackage[normalem]{ulem}
\usepackage{units}
\usepackage{dcolumn}
\usepackage{paralist}
\usepackage{natmove}
\usepackage{natbib}
\usepackage{tikz}
\usepackage{pgfplots}
\usetikzlibrary{patterns,arrows,calc,decorations.pathmorphing,shadings,shapes.geometric,positioning,fit}
\usepackage{sidecap}
\usepackage{amsmath}
\usepackage{amssymb}
\usepackage{amsfonts}
\usepackage{upgreek}
\usepackage{mathrsfs}
\usepackage{listings}
\usepackage{subfigure}
\usepackage[toc,page]{appendix}
\usepackage{siunitx}
\usepackage{comment}
\usepackage{commath}
\usepackage[breaklinks]{hyperref}

\usepackage{color}

\newcommand{\Tab}[1]{Table~\ref{tab:#1}}
\newcommand{\Equ}[1]{Eq.~\ref{equ:#1}}

\newcommand{\Fig}[1]{Fig.~\ref{fig:#1}}

\newcommand{\angs}{\unit{\AA}}

\newcommand{\avg}[1]{{\langle#1\rangle}}
\newcommand{\bavg}[1]{{\big\langle#1\big\rangle}}

\newcommand{\newtext}[1]{}

\definecolor{LighterRed}{rgb}{1.0, 0.2, 0.05}
\definecolor{ashgrey}{rgb}{0.7, 0.75, 0.71}

\begin{document}


\title{Improved general-purpose five-point model for water: TIP5P/2018}

\author{Yuriy Khalak}
\email{y.v.khalak@tue.nl}
\affiliation{Eindhoven University of Technology, Department of Mathematics and Computer
  Science \& Institute for Complex Molecular Systems, P.O. Box 513, 5600MB Eindhoven, The Netherlands}

\author{Bj\"orn Baumeier}
\email{B.Baumeier@tue.nl}
\affiliation{Eindhoven University of Technology, Department of Mathematics and Computer
  Science \& Institute for Complex Molecular Systems, P.O. Box 513, 5600MB Eindhoven, The Netherlands}

\author{Mikko Karttunen}
\email{mkarttu@gmail.com}
\affiliation{
Department of Chemistry, The University of Western Ontario,
1151 Richmond Street, London, Ontario, Canada N6A~3K7
}
\affiliation{
Department of Applied Mathematics, The University of Western Ontario,
1151 Richmond Street, London, Ontario, Canada N6A~5B7
}

\begin{abstract}
A new five point potential for liquid water, TIP5P/2018, is presented along with
the techniques used to derive its charges from \textit{ab initio} per-molecule electrostatic potentials in the liquid phase
using the split charge equilibration (SQE) of Nistor \textit{et al.} [J. Chem. Phys. \textbf{125}, 094108 (2006)].
By taking the density and diffusion dependence on temperature as target properties,
significant improvements to the behavior of isothermal compressibility were achieved along with improvements
to other thermodynamic and rotational properties. While exhibiting a dipole moment close to \textit{ab initio} values,
TIP5P/2018 suffers from a too small quadrupole moment due to the charge assignment procedure and results
in an overestimation of the dielectric constant.
\end{abstract}

\maketitle

\section {Introduction}

Water is characterized by chemical simplicity and complex microscopic and macroscopic behavior.
The textbook example of this is the density maximum at \ang{4}C and in his 2006 review Martin Chaplin mentions water having  63 anomalies\cite{chaplinweb_v2}.
Now, 12 years later that number has increased to 74\cite{Chaplin2006a}.  It is quite stunning indeed since  water is the most studied single substance.

The above alone is enough to state that modeling  water and its interactions with other molecules is a challenge.
This is well manifested in the great number of water models: In 2002, Guillot listed 46 water models of which over 30
are classical\cite{Guillot2002a}. Since then, tens of new models and tens of refinements of old ones have been introduced,
see e.g. Ref.~\citenum{Onufriev2017}. The emergence of coarse-grained and special purpose models has brought even more models
to the market including some rather unwaterlike names such as Mercedes Benz\cite{Truskett2002,Dias2009a}, BMW\cite{Wu2010a} and mW\cite{Molinero2009}.
Even non-conformal models
that do not obey the energy and distance scaling in a Lennard-Jones manner have been introduced\cite{del_Rio_1998,Rodriguez-Lopez2017}.

Given the number of models, it is perhaps somewhat surprising that most modern two-body interaction water models are based on
the functional form of the 1933 model of Bernal and Fowler\cite{Bernal1933}, that is,
the energy of two interacting water molecules is given as
\begin{equation}
E = \sum_\mathrm{pairs}
k \frac{q_i q_j}{r_{ij}} +
4 \epsilon \left[
\left(
  \frac{\sigma}{r_\mathrm{OO}}
\right)^{12}
-
\left(
  \frac{\sigma}{r_\mathrm{OO}}
  \right)^{6}
\right]
\label{eq:bernal}
\end{equation}
where $r_{ij}$ is the distance between the charges $q_i$ and $q_j$,
$k$ is the constant in Coulomb's law (containing the dielectric constant),
$r_\mathrm{OO}$ is the oxygen-oxygen distance, $\sigma$ is the distance between the oxygens at zero potential and
$\epsilon$ is the depth of the potential well. Interestingly, the most important early
computational water models, such as BNS\cite{Rahman1971} and ST2\cite{Stillinger1974}, were 5-point models.
Three-point and four-point models were introduced later to reduce the computational cost and to make the models
compatible with biomolecular force fields. It took almost 30 years from the BNS model for the general purpose 5-point model,
the TIP5P\cite{mahoney_five-site_2000}, to be introduced. The TIP5P model and its improvement are our focus in this article.

The aim of the original TIP5P model, introduced by Mahoney and Jorgensen in 2000\cite{mahoney_five-site_2000}, was to improve
on the rather poor behavior of the TIP3P\cite{jorgensen_comparison_1983} and TIP4P\cite{jorgensen_comparison_1983} models in reproducing the liquid density behavior
while keeping the model still compatible with the commonly used biomolecular force fields.
The TIP5P model was originally parameterized for \SI{9}{\angs} cutoff and reparameterized a few years later by Rick (called TIP5P-E)
for use with Ewald summation methods\cite{rick_reoptimization_2004}.
The reparameterization involved only Lennard-Jones parameters while keeping the rest of them and geometry unchanged.
In this work, we provide a new parameterization of the TIP5P model. The resulting new model is called TIP5P/2018.
Details are discussed below but in brief: 1) the geometry of the original TIP5P is retained and 2) both the charges and Lennard-Jones parameters
are modified leading to significantly improved properties.
Instead of the traditional methods, charge assignment is done using the so-called split charge equilibration (SQE) method
originally introduced by Nistor \textit{et al.}\cite{nistor_generalization_2006,nistor_dielectric_2009}.
To assign charges, we fit parameters of the SQE energy expression to per-molecule \textit{ab initio} electrostatic potentials
and use the average charges predicted by SQE. This will be detailed below.

We chose the TIP5P model as the basis due to its good basic properties and since it has a lot of potential for improvement.
It should be mentioned, however, that at this time TIP3P\cite{jorgensen_comparison_1983} is probably the most commonly used water model.
The reparameterized TIP4P model, TIP4P/2005\cite{abascal_general_2005} has gained popularity and is often
quoted as the best of the current non-polarizable general-purpose models while TIP5P has not performed up to the initial expectations,
see for example the ranking of water models by Vega and Abascal\cite{Vega2011} and other recent comparisons\cite{Agarwal2011,Fugel2017}.
As the results here show, the new TIP5P/2018 easily outperforms the previous TIP5P and TIP5P-E and compares very favorably to TIP4P/2005.
This is shown by examining a number of thermodynamic variables over a broad temperature scale.

\begin{figure}[h]
  \includegraphics[width=.8\columnwidth]{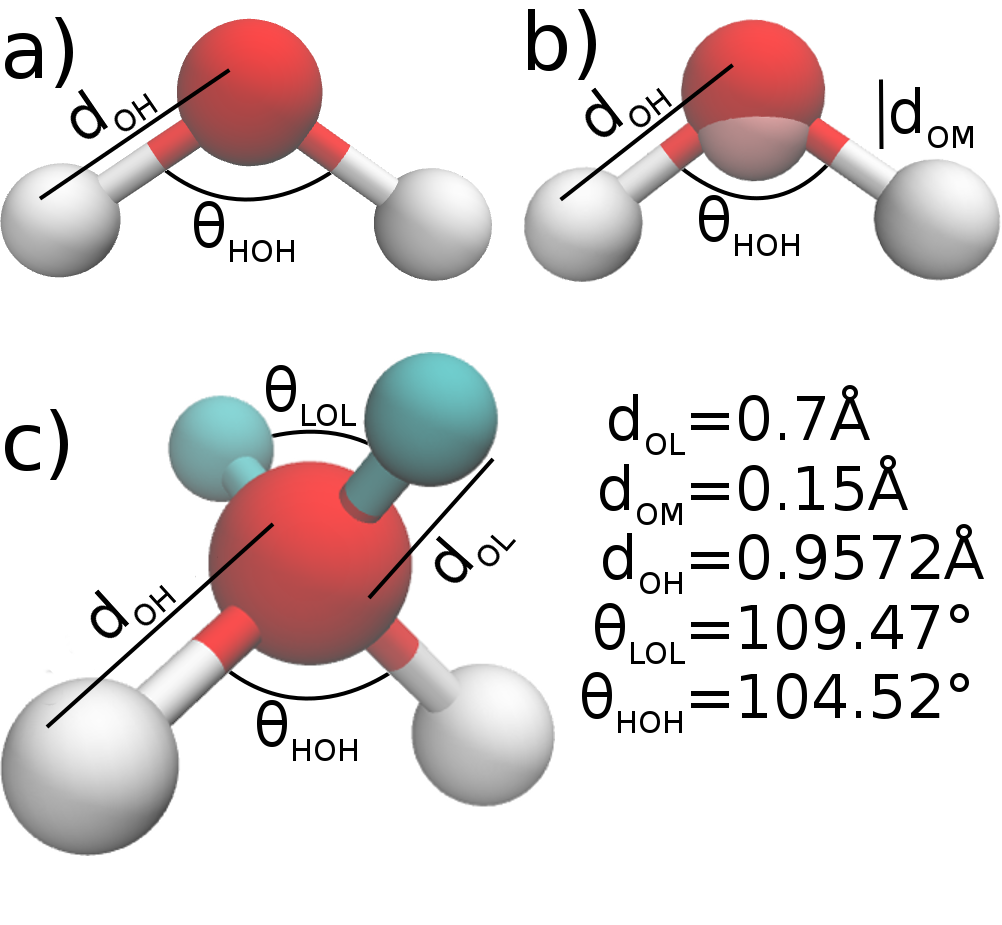}
  \caption{Geometries of TIP3P, TIP4P and TIP5P water models. a) Three point (3P) geometries only include positions of the oxygen and hydrogens.
  b) Four point (4P) geometries move the negative charge from the oxygen to a virtual site closer to the hydrogens.
  c) Five point (5P) geometries instead have two virtual sites to represent the oxygen's Lewis pairs.
  In contrast to TIP5P\cite{mahoney_five-site_2000} and TIP5P-E\cite{rick_reoptimization_2004}, TIP5P/2018 assigns
  charges to all the available five points.}
  \label{fig:geom}
\end{figure}

Like TIP3P and TIP4P/2005, TIP5P can be used in connection with many of the Amber and CHARMM force field variants for bimolecular simulations, see Fig.~\ref{fig:geom} for the geometries.
It has, however, not become widely used and has received mixed reviews\cite{Nutt2007,GlassKrishnanNuttEtAl2010,Florova2010,Kuehrova2013}.
The notable exception is the recent comparison of polysaccharide force fields by Sauter and Grafm{\"u}ller who wrote
\textit{''we conclude that GLYCAM TIP5P is best suited for studying oligosaccharides''}\cite{Sauter2015}.
Based on the bulk properties of the new TIP5P/2018 model presented here, we expect good performance in connection with biomolecular force fields. Testing that, however, is beyond the current study.

\section {Methods}

\newcommand{\Qt}[1]{{\widetilde{Q}_{#1}}}

\subsection{Initial Charge Assignment} \label{ss:met_charge_fitting}

Electrostatic potential (ESP)
fitting\cite{Momany1978,Cox1981,Singh1984} is a classical approach to
assigning charges for use in molecular dynamics (MD) potentials. It
involves minimizing the difference between an \textit{ab initio}
reference electrostatic potential and that produced by the assigned
charges on a grid within a region close the molecule.  Grid points too
close to the nuclei are excluded, as the electron density there has
too much structure to be accurately represented by point
charges. Similarly, grid points at large distances (typically
$>\SI{3}{\AA}$) are not taken into account for computational
efficiency. The exact definition of the grid is
method-specific\cite{breneman_determining_1990,Singh1984}. One of known problems of
these methods is that  atoms embedded at the centers of larger molecules are assigned
charges with values that are not chemically intuitive, mostly due to
the lack of grid points nearby these atoms.

Restrained electrostatic potential (RESP)
fitting\cite{bayly_well-behaved_1993,cornell_application_1993}
combats this by imposing a harmonic charge restraint of an \textit{a
  priori} assigned weight preventing the assigned charges from
significantly deviating from predetermined values, typically
zeros. RESP has become popular for building force
fields\cite{cieplak_application_1995,wang_development_2004,dupradeau_r.e.d._2010},
however in bulk water all the atoms effectively become embedded and
have few surrounding grid points.
Furthermore, these points are located in the empty space between the
molecules. In molecular dynamics electrostatic forces act between
pairs of nuclei, meaning that such a grid samples regions with low
importance for parameterization. In this situation, the restraint of
RESP plays a disproportionate role in charge assignment.

To avoid the above situation, we use a different approach. We
determine the bulk wave function of a periodic 54 water molecule
structure by Titantah and
Karttunen\cite{titantah_hydrophobicity_2015-1} using CPMD 3.17.1\cite{CPMD_v2}, the BLYP
functional\cite{becke_density-functional_1988-1,lee_development_1988-1}
with D3 van der Waals
corrections\cite{grimme_consistent_2010,hujo_performance_2013},
pseudopotentials in Troullier-Martins form for oxygen\cite{troullier_efficient_1991} and
Kleinman-Bylander form\cite{kleinman_efficacious_1982} for hydrogen, and a \SI{90}{Ry} plane
wave energy cutoff. This wave function is then projected onto a
Gaussian atom-centered basis. For this, we use the basis set for the
Stuttgart/Dresden pseudopotentials\cite{bergner_ab_1993},
which are augmented, after uncontraction, by additional polarization
functions\cite{krishnan_self-consistent_1980} of $d$ symmetry. This
results in a projection completeness above 99.95\%.
We then subdivide the overall electron density into individual
molecular contributions, which are determined from the full
atomic-orbital density matrix after projection, by assigning its rows
to the respective molecules. This ensures that electron density
resulting from overlap of basis functions from two different
molecules, described by off-diagonal blocks in the density matrix, is
equally split among them.

The above subdivision allows one to compute molecular electrostatic
potentials based on per-molecule electron densities that include the
effects of polarization from other molecules in the bulk. At the same
time, no grid points are excluded by the presence of other molecules
in the setup of the grid for ESP fitting, permitting to sample the
most relevant regions of the electrostatic potential for molecular
dynamics. We implemented this procedure as a branch\footnote{Available
from \url{https://github.com/votca/xtp/tree/periodic_integration}}
of the open source VOTCA-XTP package\cite{XTP_v2}
in a manner that includes Ewald summation for periodic systems\cite{ewald_berechnung_1921-1}.

Next, we turn to the SQE
formalism\cite{nistor_generalization_2006,nistor_dielectric_2009} as
a restraint-free alternative to RESP for charge assignment.  SQE is a
recent iteration on electronegativity and charge equilibration
methods\cite{mortier_electronegativity_1985,rappe_charge_1991,rick_dynamical_1994}.
In contrast to earlier methods of this type, SQE combines both atom
and bond based energy terms to enforce charge neutrality of
interacting closed shell molecules without the need for charge
restraints. It also introduces a penalty to long-range charge
transfer, preventing long chain molecules from being overly polarizable, a
problem common in the older techniques\cite{lee_warren_origin_2008}.
SQE has been successfully used to describe redox reactions in
batteries\cite{dapp_towards_2013,dapp_redox_2013} and has been shown
to reproduce the behavior of charges and electrostatic potentials in
applied electric fields\cite{verstraelen_computation_2012}.
Specifically, we rely on the SQE energy expression
\begin{equation}\label{equ:SQE_U}
U = \chi_i Q_{i} + \frac{1}{2} K_i Q_{i}^2 +
\frac{1}{2}\sum_i{\sum_j{K^{\mathrm{bond}}_{ij}(r_{ij})q_{ij}^2}} +
U_c \,,
\end{equation}
where $Q_i=\sum_j{q_{ij}}$ is the atomic charge assigned to atom $i$
as a result of contributions $q_{ij}=-q_{ji}$ from its covalently
bonded neighbors $j$. The electronic parameters $\chi_i$ and $K_i$
represent electronegativity and atomic hardness, respectively.
The distance-dependent bond hardness
$K^{\mathrm{bond}}_{ij}(r_{ij}) = \exp{(r_{ij}/\alpha_{ij})} -1$
acts as the penalty for charge
transfer exceeding distances characterized by $\alpha_{ij}$.
The energy due to Coulomb interactions $U_c$ between the charges
includes the effects of periodicity, through Ewald
summation\cite{ewald_berechnung_1921-1}, and intramolecular shielding (see
Appendix).

For any set of the SQE parameters, atomic charges $Q_i$ can be obtained
by minimizing~\Equ{SQE_U} with respect to the charge transfers
$q_{ij}$ along covalent bonds. We then apply the ESP procedure to
iteratively optimize the SQE parameters and obtain a distribution of
charges for each site. These distributions are very narrow with
standard deviations of less than $0.014$ elementary charges.
We use their means as the charges on a rigid five point (\Fig{geom}c)
geometry in further potential refinement.

\subsection{Parameter optimization} \label{ss:met_param_opt}

The charges obtained using the procedure above are combined with
Lennard-Jones parameters fitted to reproduce the experimental
dependence of density on temperature, especially the density peak at
\SI{4}{\celsius}.  At this stage, without further adjustments, the
resulting potential yields significantly smaller diffusion
than experiments ($1.60\times10^{-5}$ compared to \SI{2.30e-5}{cm^2/s}~\cite{pruppacher_selfdiffusion_1972}).
In addition, the strength of the hard
core repulsion required to correctly position the density peak is much
larger than that of other water models.
Besides the Lennard-Jones interaction, the only other
interactions in the system are electrostatic, see
Eq.~\ref{eq:bernal}. Therefore, to increase diffusion while
maintaining the position of the density peak we explore  scaling of
the charges.

Uniform charge scaling has previously been suggested to correct ion
binding strength for the CHARMM and AMBER
forcefields\cite{leontyev_accounting_2011}.  While integer ion
charges represent the reality in vacuum, once inserted into water, the
effective charge felt by the surrounding molecules is smaller due to
screening.  Non-polarizable water models can account for some of this
effect explicitly by molecular rearrangement. However, without a
polarizable description, screening due to changes in electron density
are not explicitly accounted for.  The electronic continuum
correction\cite{leontyev_electronic_2010,leontyev_accounting_2011,leontyev_polarizable_2012,pluharova_ion_2013,vazdar_aqueous_2013,kohagen_accurate_2014,kann_scaled-ionic-charge_2014,kohagen_accounting_2016}
provides a way to implicitly model this effect by embedding the system
in a continuous dielectric medium with a dielectric constant
$\epsilon_\mathrm{el}$. This is equivalent to scaling the ionic charge $Q$ such
that the effective charge is
$Q_\mathrm{eff}=Q/\sqrt{\epsilon_\mathrm{el}}$. For liquid water the
scaling factor is about 0.75\cite{leontyev_accounting_2011} and the method has been successfully
applied in simulations of biomolecular systems\cite{melcr_accurate_2018,duboue-dijon_binding_2018}.

The formulation of SQE we used in this work already includes the intramolecular electronic screening contributions through shielded electrostatic interactions (see Appendix). Intermolecular contributions, though, were not explicitly handled. Therefore, some charge scaling was still necessary.
By scaling all charges by 0.95 and once again reoptimizing the Lennard-Jones parameters similar to Refs.~\onlinecite{kohagen_accurate_2014,kohagen_accounting_2016}, we are able to recover the correct self-diffusion behavior.  This is the parameterization of TIP5P/2018 as is detailed below in Sec.~\ref{ss:res_dens_diff}.

All three TIP5P-based models have the same geometry, that is,
an oxygen with the sole Lennard-Jones interaction site, two hydrogens \SI{0.9572}{\angs}
from the oxygen and separated by a \SI{104.52}{\degree} angle, and two virtual sites
\SI{0.7}{\angs} from the oxygen separated by a perfect tetrahedral angle of \SI{109.47}{\degree}, \Fig{geom}c.
The main difference between TIP5P/2018 and the previous TIP5P models
is that charges are assigned to every site.
The large portion of the charge that in the case of TIP5P is located on
the virtual sites is instead on the oxygen. This reduces tetrahedrality (Sec.~\ref{ss:res_tetrahedral_order}),
a problem TIP5P has been criticized\cite{Vega2011}.
The exact parameters of the TIP5P/2018 potential are presented in Table~\ref{tab:params}.

\begin{table}[]
\begin{tabular}{lccc}
                                                   & TIP5P\cite{mahoney_five-site_2000}
                                                             & TIP5P-E\cite{rick_reoptimization_2004}
                                                                       & TIP5P/2018
                                                                                   \\ \hline
\multicolumn{1}{l|}{$Q_{O}$}                       & 0       & 0       &-0.641114  \\
\multicolumn{1}{l|}{$Q_{H}$}                       & 0.241   & 0.241   & 0.394137  \\
\multicolumn{1}{l|}{$Q_{L}$}                       &-0.241   &-0.241   &-0.073580  \\
\multicolumn{1}{l|}{$\epsilon$ (kJ/mol)}           & 0.66944 & 0.744752& 0.79      \\
\multicolumn{1}{l|}{$\sigma$ (\angs)}              & 3.12    & 3.097   & 3.145     \\
\end{tabular}
\caption {Parameters defining five point models:
site charges $Q$, Lennard-Jones potential well depth $\epsilon$, and distance of zero potential $\sigma$.
All these five point potentials share the same geometry, \Fig{geom}c.}
\label{tab:params}
\end{table}

\subsection{Computational details} \label{ss:ComputDetails}

The Gromacs 2016.3\cite{abraham_gromacs_2015} software package
was used for all MD simulations.
All TIP5P/2018 simulations use a time step of \SI{2}{\femto\second},
the smooth particle mesh Ewald method\cite{essmann_smooth_1995-1} (SPME) for
computing the electrostatic interactions and \SI{0.85}{\nano\meter} cutoff for both the Lennard-Jones
and real-space part of the Coulomb interactions.
This cutoff is smaller than those of TIP5P and TIP5P-E, and has been adopted from TIP4P/2005 in an attempt to reduce the computational expense.
Dispersion corrections for energy and pressure were also used.
Unless stated otherwise,
all thermostat and barostat time constants were set to \SI{1}{\pico\second} and
all target pressures were set to \SI{1}{atm}.

Thermodynamic properties across the range
\SIrange{235.65}{348.15}{\kelvin} were obtained by averaging over time and five  \SI{20}{\nano\second} replicas at each temperature.
Equilibration for each replica was performed as follows:
First, a cubic simulation box with a side length of \SI{4}{\nano\meter} containing 2,069 water molecules was set up and  energy was minimized with the steepest descent algorithm. Next,
a \SI{10}{\pico\second} simulation under the
canonical ensemble with the Berendsen thermostat\cite{berendsen_molecular_1984} (with time constant $\tau_t=\SI{0.5}{\pico\second}$)
followed by a \SI{20}{\pico\second}
isothermal-isobaric simulation with the Nos\'e-Hoover thermostat\cite{nose_molecular_1984,hoover_canonical_1985-1}
($\tau_t=\SI{1}{\pico\second}$) and Berendsen barostat\cite{berendsen_molecular_1984}
(at \SI{1}{atm} with time constant $\tau_p=\SI{1}{\pico\second}$) was performed.
After equilibration, \SI{20}{\nano\second} production simulation
with Nos\'e-Hoover thermostat\cite{nose_molecular_1984,hoover_canonical_1985-1} ($\tau_t=\SI{1}{\pico\second}$)
and Parrinello-Rahman barostat\cite{parrinello_crystal_1980,parrinello_polymorphic_1981-1} (at \SI{1}{atm}, $\tau_p=\SI{1}{\pico\second}$) followed.
The first nanosecond of the production
simulations was excluded from analysis as final equilibration.

Diffusion was computed from mean square displacement,
while the thermodynamic properties were computed from drift corrected fluctuations of volume, density, and potential energy using standard formulas\cite{allen_computer_1987}.

For comparison, we also performed control simulations with TIP4P/2005, TIP5P, and TIP5P-E.
The TIP4P/2005 simulations were carried out under the same conditions as TIP5P/2018, except with \SI{9}{\angs} cutoffs.
When simulating TIP5P and TIP5P-E,
we used \SI{9}{\angs} cutoffs, 512 molecule systems, a \SI{1}{fs} time step, and no dispersion
corrections, to match conditions under which they were
parameterized\cite{mahoney_diffusion_2000,rick_reoptimization_2004}.
To enforce densities originally reported for TIP5P and TIP5P-E, we used the canonical ensemble for both of these models.
For TIP5P, we used the Berendsen thermostat\cite{berendsen_molecular_1984} with $\tau_p=\SI{0.1}{\pico\second}$ and simple cutoffs,
while for TIP5P-E we used the Nos\'e-Hoover thermostat\cite{nose_molecular_1984,hoover_canonical_1985-1}
with $\tau_p=\SI{1}{\pico\second}$ and SPME.
Wherever possible, we compare properties of TIP5P/2018 to those of TIP5P and
TIP5P-E reported by their original authors. 
Comparisons with experimental data are provided when possible.

\section {Results}

The properties of TIP5P/2018 are summarized in Table~\ref{tab:results} while their temperature dependencies
are discussed in the sections below.

\begin{table*}[]
\begin{tabular}{lcccccccc}
                                                   & $\rho$   & $D$ & $\alpha_P$ & $\kappa$ & $C_P$ & $\epsilon$ & $\tau_1^\mathrm{dipole\,long}$ & $\tau_2^\mathrm{HH\,long}$\\
                                                   & (\si{\kilo\gram\per\metre\cubed})   & ($10^{-5}\times$\si{\centi\metre\squared\per\second})
                                                    & ($10^{5}\times$\si{\per\kelvin}) & ($10^{6}\times$\si{atm^{-1}})
                                                     & (\si{\J\per\mole\per\kelvin})& & (\si{\pico\second}) & (\si{\pico\second})\\
                                                   \hline
TIP5P
                                                   & $999\pm1$\cite{mahoney_five-site_2000}          & $2.62\pm0.04$\cite{mahoney_diffusion_2000}     & $63\pm6$\cite{mahoney_five-site_2000}   & $41\pm2$\cite{mahoney_five-site_2000}     & $122\pm3$\cite{mahoney_five-site_2000}    & $82\pm2$\cite{rick_reoptimization_2004}  & $5.788$ & $2.506$ \\
TIP5P-E
                                                   & $1000.0\pm0.5$\cite{rick_reoptimization_2004}   & $2.8\pm1$\cite{rick_reoptimization_2004}           & $49\pm6$\cite{rick_reoptimization_2004}  & $52\pm3$\cite{rick_reoptimization_2004}   & $114\pm3$\cite{rick_reoptimization_2004}  & $92\pm14$\cite{rick_reoptimization_2004} & $5.521$ & $2.455$ \\
TIP5P/2018
                                                   & $996.30\pm0.06$                                  & $2.34\pm0.02$                                       & $42\pm3$                                  & $48.2\pm0.5$                               & $105\pm4$                                  & $127\pm4$                                 & $6.575$ & $2.901$ \\
experimental
                                                   & $997.05$\cite{Haynes2025crc}                    & $2.30$\cite{pruppacher_selfdiffusion_1972}         & $25.7$\cite{kell_density_1975}   & $45.85$\cite{kell_density_1975}           & $75.36$\cite{kell_notitle_1972}             & $78.3$\cite{malmberg_dielectric_1956}    & $\approx 7.69$\cite{ronne_investigation_1997}    & $\approx 2.46$\cite{jonas_molecular_1976} \\

\end{tabular}
\caption {Properties of TIP5P, TIP5P-E, and TIP5P/2018 at \SI{298.15}{\kelvin} and \SI{1}{atm}.
  Experimental density\cite{Haynes2025crc} is reported at \SI{1}{bar}.
  The experimental values for $\tau_1^\mathrm{dipole\,long}$ and $\tau_2^\mathrm{HH\,long}$
  are estimates linearly interpolated from measurements\cite{ronne_investigation_1997,jonas_molecular_1976}
  at nearby temperatures.
}
\label{tab:results}
\end{table*}

\subsection{Density and Diffusion} \label{ss:res_dens_diff}
\begin{figure}
    \includegraphics[width=\columnwidth]{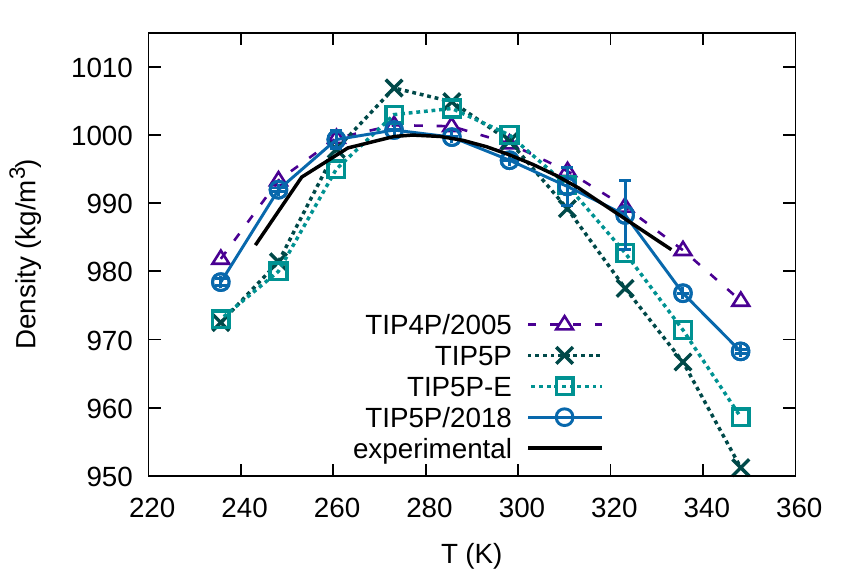}
    \caption{Density as a function of temperature at a pressure of \SI{1}{atm}.
	     TIP5P and TIP5P-E values are obtained from Ref. \onlinecite{rick_reoptimization_2004}.
	     TIP5P/2018 and TIP4P/2005 values are results of our own simulations.
	     Error bars for TIP5P/2018 represent standard deviations of the results from five simulations.
	     Experimental data\cite{Haynes2025crc,hare_density_1987}
	     is shown for comparison}
    \label{fig:thermo_rho}
\end{figure}

Both the maximum density $\rho_\mathrm{max}$ and the corresponding temperature $T_\mathrm{max}$ for TIP5P/2018
match those of experiments\cite{Haynes2025crc,hare_density_1987} (\Fig{thermo_rho}).
This is a marked improvement over both TIP5P and TIP5P-E, which, while capturing the correct
$T_\mathrm{max}$, exhibit an overly high $\rho_\mathrm{max}$ and a density that decays too rapidly. 
TIP5P/2018 significantly improves on the density decay compared to TIP5P and TIP5P-E, but is still unable
to fully reproduce the experimental profile at high temperatures. Attempts to further flatten the density
profile during potential optimization led to significantly lower diffusion accompanied by more first solvation shell structuring.

\begin{figure}
    \includegraphics[width=\columnwidth]{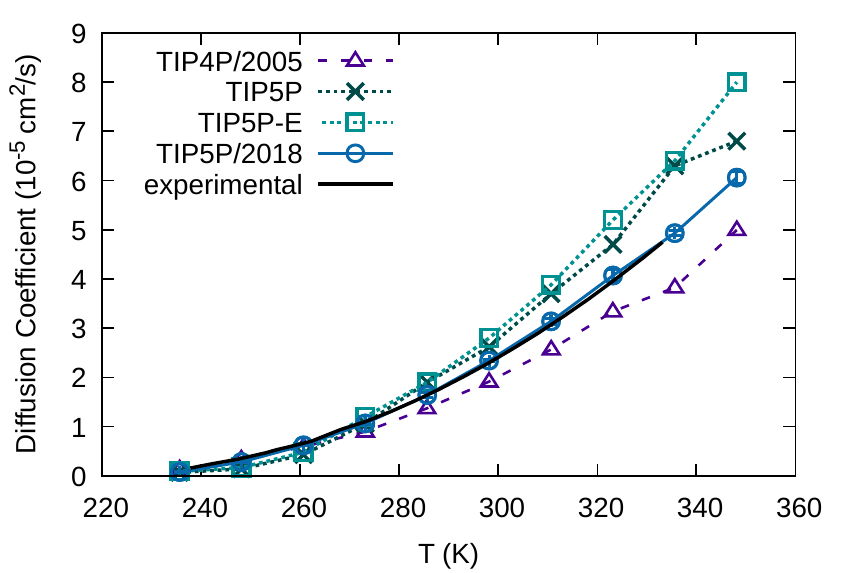}
    \caption{Self-diffusion as a function of temperature at a pressure of \SI{1}{atm}.
	     TIP5P and TIP5P-E values are obtained from Refs. \onlinecite{mahoney_five-site_2000,rick_reoptimization_2004}
	     and the experimental values from Refs. \onlinecite{holz_temperature-dependent_2000,price_self-diffusion_1999}.
		 TIP5P/2018 and TIP4P/2005 values are results of our own simulations.
	     Error bars for TIP5P/2018 represent standard deviations of the results from five simulations.}
    \label{fig:thermo_diff}
\end{figure}

On the other hand, as Fig.~\ref{fig:thermo_diff} shows, TIP5P/2018 reproduces the diffusion profile extremely well,
which is not surprising as it was one of the target properties during optimization. In comparison, TIP5P
and TIP5P-E both exhibit much faster diffusion at higher temperatures while underestimating the experimental
diffusion values at and below \SI{262.65}{\kelvin}. TIP4P/2005, meanwhile, underestimates diffusion at higher temperatures.

\subsection{Radial Distribution Functions} \label{ss:res_rdf}

\begin{figure}
    \includegraphics[width=\columnwidth]{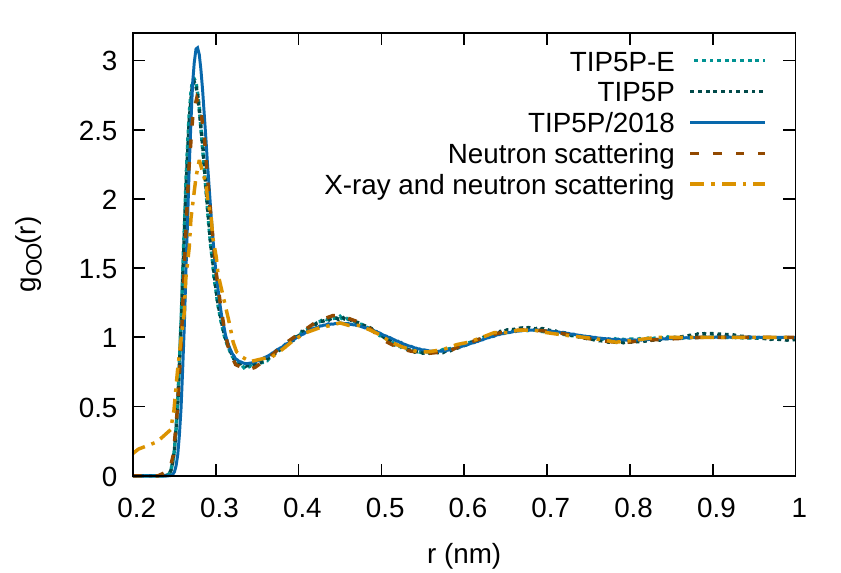}
    \caption{Oxygen-oxygen radial distribution functions for TIP5P/2018 (\SI{298}{\kelvin}), TIP5P (\SI{298.15}{\kelvin}),
    and TIP5P-E (\SI{298.15}{\kelvin}). Data for all these models comes from our own simulations.
    Experimental radial distribution functions from neutron scattering (\SI{298}{\kelvin})\cite{soper_radial_2000}
    and combined X-ray and neutron scattering (\SI{300}{\kelvin})\cite{soper_radial_2013} are also shown.}
    \label{fig:rdf_OO}
\end{figure}

\begin{figure}
    \includegraphics[width=\columnwidth]{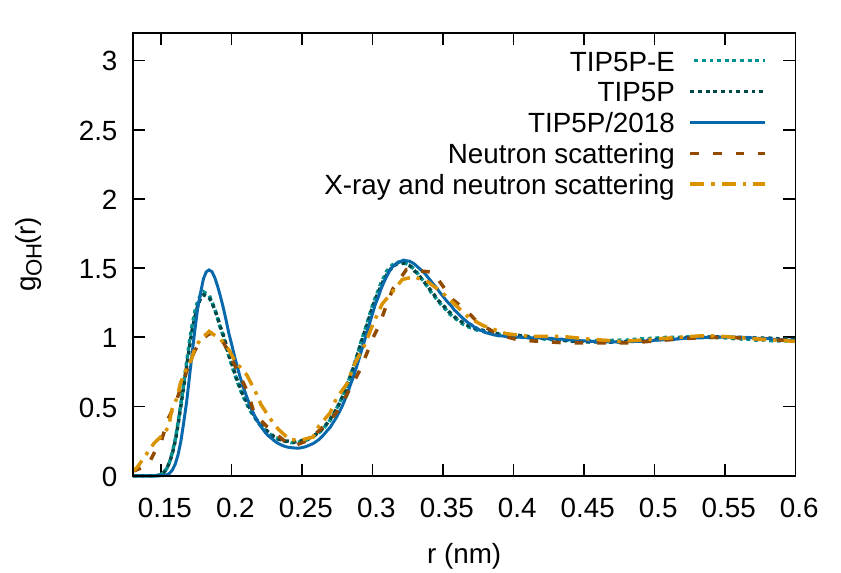}
    \caption{Oxygen-hydrogen radial distribution functions for TIP5P/2018 (\SI{298}{\kelvin}), TIP5P (\SI{298.15}{\kelvin}),
    and TIP5P-E (\SI{298.15}{\kelvin}). Data for all these models comes from our own simulations.
    Experimental radial distribution functions from neutron scattering (\SI{298}{\kelvin})\cite{soper_radial_2000}
    and combined X-ray and neutron scattering (\SI{300}{\kelvin})\cite{soper_radial_2013} are also shown.}
    \label{fig:rdf_OH}
\end{figure}

\begin{figure}
    \includegraphics[width=\columnwidth]{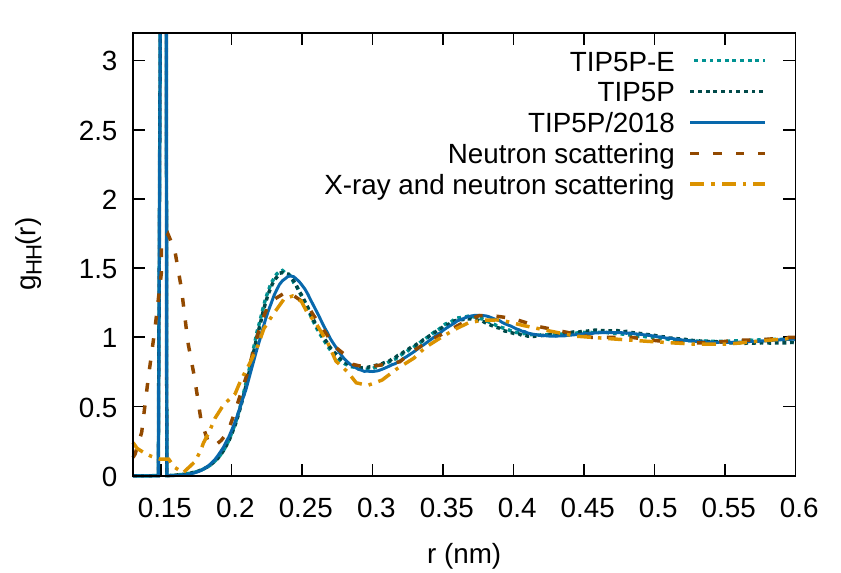}
    \caption{Hydrogen-hydrogen radial distribution functions for TIP5P/2018 (\SI{298}{\kelvin}), TIP5P (\SI{298.15}{\kelvin}),
    and TIP5P-E (\SI{298.15}{\kelvin}). Data for all these models comes from our own simulations.
    Experimental radial distribution functions from neutron scattering (\SI{298}{\kelvin})\cite{soper_radial_2000}
    and combined X-ray and neutron scattering (\SI{300}{\kelvin})\cite{soper_radial_2013} are also shown.
    The sharp peak near \SI{1.5139}{\angs} corresponds to the intramolecular hydrogen-hydrogen distance and is the result
    of using a rigid geometry.}
    \label{fig:rdf_HH}
\end{figure}

The radial distribution functions of TIP5P/2018 are in good agreement with experiments (Figs.~\ref{fig:rdf_OO}, \ref{fig:rdf_OH} and \ref{fig:rdf_HH}).
The major discrepancy is the  higher first peak of the oxygen-oxygen radial density function, Fig.~\ref{fig:rdf_OO}.
Additionally, neutron scattering experiments show a wide first peak for $g_\mathrm{HH}(r)$ (Fig.~\ref{fig:rdf_HH}), which corresponds to the intramolecular H-H distance.
Due to the rigidity of the five point potentials this distance is constrained to
\SI{1.5139}{\angs} in the simulations, and appears as a sharp peak.

Rigidity of the five-point geometry also contributes to the first intermolecular $g_\mathrm{OH}(r)$ peak (Fig.~\ref{fig:rdf_OH}), corresponding to the hydrogen bonding interaction.
In all the five point models this peak is narrower and taller than in experiments. Rigid O-H covalent bonds in the simulations lead to reduced smearing of this peak from the optimal hydrogen bonding distance
than is present in real water. 
Because of these effects, it is reasonable to assume that using a flexible geometry would better describe liquid water.
However, a flexible model without explicit treatment of polarization tends to result in incorrect dipole moment dependence on geometry as discussed in Ref. \onlinecite{mahoney_five-site_2000}.

\subsection{Orientational Tetrahedral Order}\label{ss:res_tetrahedral_order}

The orientational tetrahedral order $q$ is a local measure of the angular alignment indicating how well the surroundings of a
water molecule reproduce the regular tetrahedral structure (although slightly inconvenient here, we are using the notation of the original paper with $q$ denoting the tetrahedral order parameter). The expression we employ for $q$
was originally proposed by Chau and Hardwick\cite{chau_new_1998}
and rescaled by Errington and Debenedetti\cite{errington_relationship_2001} to produce values between 0 and 1,
\begin{equation} \label{equ:tetrahedral_order}
q = 1 - \frac{3}{8} \sum_{j=1}^3 \sum_{k=j+1}^4 \left( \cos \phi_{jk} + \frac{1}{3} \right)^2 \,,
\end{equation}
where for any given water molecule indexes $j$ and $k$ iterate over the oxygens of its four nearest neighbors and
$\phi_{jk}$ is the angle between these neighbors centered on the oxygen of the water molecule in question.
A perfect tetrahedral arrangement, similar to that of hexagonal ice ($\text{I}_\text{h}$), occurs at $q=1$,
while an ideal gas corresponds to $q=0$.
Orientational tetrahedral order has previously been used to study both supercooled water and water in proximity
to various solutes, as cited in Ref. \onlinecite{duboue-dijon_characterization_2015}. For liquid water, distribution $f(q)$ typically exhibits
two peaks corresponding to an ice-like population at high $q$ and a more disordered population at lower $q$.
As temperatures grow, the more disordered population becomes preferred and the whole distribution shifts to lower
values of $q$\cite{overduin_understanding_2012,titantah_hydrophobicity_2015}.
TIP5P/2018 possesses the same behavior, however the high $q$ peak is shifted to the left compared to \textit{ab initio}
results\cite{titantah_hydrophobicity_2015}, a feature shared with the TIP4P/2005 potential\cite{overduin_understanding_2012}.

\begin{figure}
    \includegraphics[width=\columnwidth]{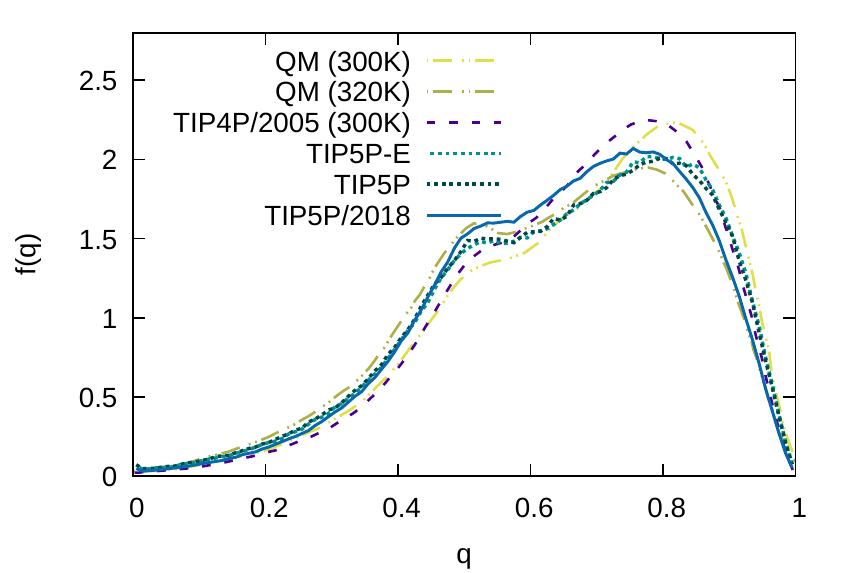}
    \caption{Normalized distributions of the orientational tetrahedral order $q$ for TIP5P/2018, TIP5P, and TIP5P-E at \SI{298}{\kelvin},
    as well as TIP4P/2005 at \SI{300}{\kelvin}\cite{overduin_understanding_2012} and \textit{ab initio} distributions
    at \SI{300}{\kelvin} and \SI{320}{\kelvin}\cite{titantah_hydrophobicity_2015}.
    Data for all five point models is from our own simulations.
    TIP5P/2018 has angular structure akin to that of other potentials at higher temperature.}
    \label{fig:tetrahedral_order}
\end{figure}

Additionally, TIP5P/2018 exhibits a larger disordered population than other models.
The ratio of peak heights resembles that of
other models at a higher temperature (\Fig{tetrahedral_order}). The simplest plausible explanation for this is too weak hydrogen bonding.
This can be illustrated with the early version of our potential without charge scaling and, therefore, stronger hydrogen bonding.
The corresponding $f(q)$ peaks of the earlier version are at the the same values of $q$, but with a larger preference for the ice-like peak, almost reaching that
of TIP4P/2005\cite{overduin_understanding_2012} and \textit{ab initio} results\cite{titantah_hydrophobicity_2015}.
Unfortunately, the earlier versions of TIP5P/2018 exhibited drastically lower diffusion as a result of the increased
hydrogen bonding and were deemed too inaccurate.

\subsection{Dielectric Constant}\label{ss:res_diel}

The dielectric constant, also known as the relative permittivity $\epsilon$ of a material,
is a key property for modeling the accurate solvation of ions.
In molecular dynamics, $\epsilon$ is typically calculated
from dipole moment fluctuations.
Because of this, the value for the dielectric constant takes long simulation
times to converge, especially at lower temperatures.  The functional form proposed by Neumann\cite{neumann_dipole_1983} is
\begin{equation} \label{equ:epsilon}
\epsilon = 1 + \frac{1}{3 \epsilon_0 k \avg{T} \avg{V}} (\avg{\vec{M}^2}-\avg{\vec{M}}^2) \,,
\end{equation}
where $\vec{M}$ is the total system dipole moment, $T$ is temperature, $V$ is the volume.

\begin{figure}
    \includegraphics[width=\columnwidth]{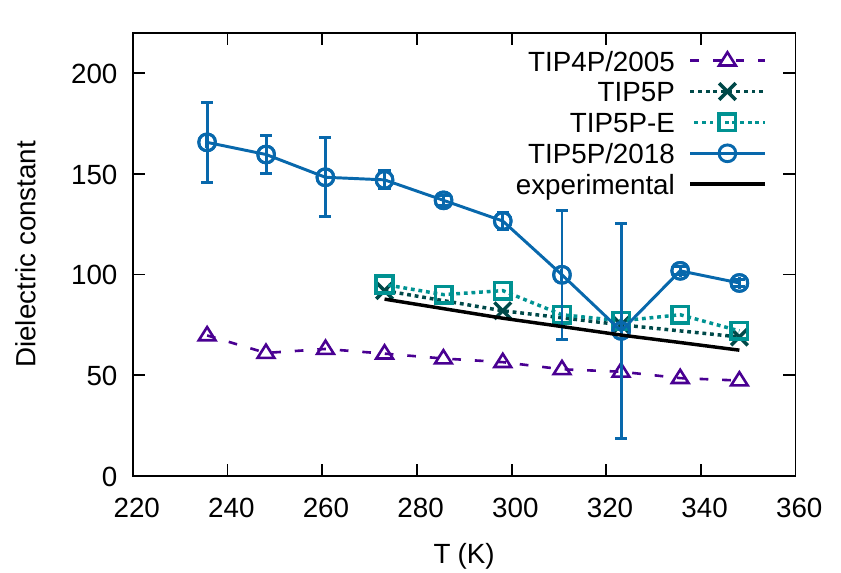}
    \caption{Dielectric constant of TIP5P/2018 as a function of temperature at a pressure of \SI{1}{atm}
	     compared to those of other models\cite{mahoney_five-site_2000,rick_reoptimization_2004}
	     and experimental values\cite{malmberg_dielectric_1956}.
	     Values for TIP4P/2005 and TIP5P/2018 are obtained from fluctuations of system dipole moment
	     in our own simulations using standard formulas\cite{allen_computer_1987}.
	     Points and error bars for TIP5P/2018 represent standard deviations and means of the results from five simulations.}
    \label{fig:thermo_epsilon}
\end{figure}

While TIP5P and TIP5P-E reproduce the experimental\cite{malmberg_dielectric_1956} temperature dependence
of this property rather well, TIP5P/2018 significantly overestimates it (\Fig{thermo_epsilon}). As explored by
Carnie and Patey\cite{carnie_fluids_1982} and later by Rick\cite{rick_reoptimization_2004},
this overestimation can have two sources: a large molecular dipole moment $\mu$,
and a small quadrupole. Both of the above works show that even at larger values of $\mu$, a large quadrupole
can quench fluctuations of the system dipole moment and lower the dielectric constant.

\begin{table}[]
\begin{tabular}{lcccc}
                                                   & $\mu$   & $Q_{xx}$& $Q_{yy}$ & $Q_{zz}$\\
                                                   & (D)     & (D \angs)& (D \angs) & (D \angs)\\
                                                   \hline
\multicolumn{1}{l|}{TIP3P\cite{jorgensen_comparison_1983}}
                                                   & 2.35    & -1.865  & 1.605  &  0.23  \\
\multicolumn{1}{l|}{TIP4P\cite{jorgensen_temperature_1985}}
                                                   & 2.18    & -2.235  & 2.065  &  0.17  \\
\multicolumn{1}{l|}{TIP4P/2005\cite{abascal_general_2005}}
                                                   & 2.305   & -2.39   & 2.21   &  0.18 \\
\multicolumn{1}{l|}{TIP5P\cite{mahoney_five-site_2000} and TIP5P-E\cite{rick_reoptimization_2004}}
                                                   & 2.29    & -1.63   & 1.50   &  0.13  \\

\multicolumn{1}{l|}{\textbf{TIP5P/2018}}           & 2.504   & -1.91   & 1.69   &  0.21 \\

\multicolumn{1}{l|}{QM surrounded by 4 TIP5P\cite{niu_large_2011}}
                                                   & 2.69    & -3.08   & 2.82   &  0.26 \\
\multicolumn{1}{l|}{QM surrounded by 230 TIP5P\cite{coutinho_electronic_2003}}
                                                   & 2.55    & -2.91   & 2.71   &  0.20 \\
\multicolumn{1}{l|}{Bulk QM (BLYP, \SI{70}{Ry})\cite{silvestrelli_structural_1999-1}}
                                                   & 2.95    & -3.36   & 3.18   &  0.18 \\
\multicolumn{1}{l|}{Bulk QM (BLYP, \SI{60}{Ry})\cite{site_electrostatic_1999}}
                                                   & 2.43    & -2.77   & 2.67   &  0.10 \\

\end{tabular}
\caption {Dipole $\mu$ and quadrupole $Q$ moments of TIP5P/2018 and similar non-polarizable potentials
as well as some \textit{ab initio} results for liquid water.
Oxygen position is taken as the point of origin. The hydrogens are located in the positive $z$ half of the $yz$ plane,
and the Lewis pairs in the $xz$ plane. Data presented for potentials other than TIP5P/2018 is derived
from the work of Niu \textit{et al.}\cite{niu_large_2011}.
The quadrupole moment definition of Stone\cite{stone_theory_2013-1} is used.}
\label{tab:multipole}
\end{table}

Even after charge scaling, TIP5P/2018 has a relatively large dipole moment
more in line with those obtained from \textit{ab initio} simulations
than with that of other rigid point charge potentials for water (\Tab{multipole}).
The quadrupole moment, on the other hand, is much smaller than in the \textit{ab initio} systems,
explaining the high dielectric constant.
This is a consequence of assigning charges based on ESP fitting with a small number of charge sites.
Reproducing both the dipole and the quadrupole moments becomes difficult in such cases\cite{niu_large_2011}.
This also illustrates why most water potentials are fully empirical by their  nature.

A possible improvement could arise from using a six-point geometry, like
that of Nada and van der Eerden\cite{nada_intermolecular_2003},
where the oxygen charge is shifted closer to the hydrogens,
similar to what occurs in four point potentials\cite{jorgensen_temperature_1985,abascal_general_2005}.
Applying such a shift to existing charges would increase the quadrupole moment
while decreasing the dipole moment.
Furthermore, the central charge site would lie further away from the Lewis pairs, allowing them to capture
more of the molecule's \textit{ab initio} charge distribution during ESP fitting
and could potentially lend the resulting potential a larger quadrupole moment.

\subsection{Thermodynamic Properties}\label{ss:res_thermo}

The majority of the thermodynamic properties of TIP5P/2018 are in better qualitative agreement with
experiments over a wider temperature range than the other five-point models are.
The temperature dependence of the coefficient of thermal expansion $\alpha_P$
for TIP5P/2018 has the same shape as that of real water, but is slightly shifted 
toward higher values.
While the corresponding profiles of both TIP5P and TIP5P-E cross the experimental profile
near \SI{280}{\kelvin}, the TIP5P $\alpha_P$ changes significantly faster than
in experiments and the TIP5P-E profile crosses the experimental line for a second time and
returns back to positive values at low temperatures.

\begin{figure}
    \includegraphics[width=\columnwidth]{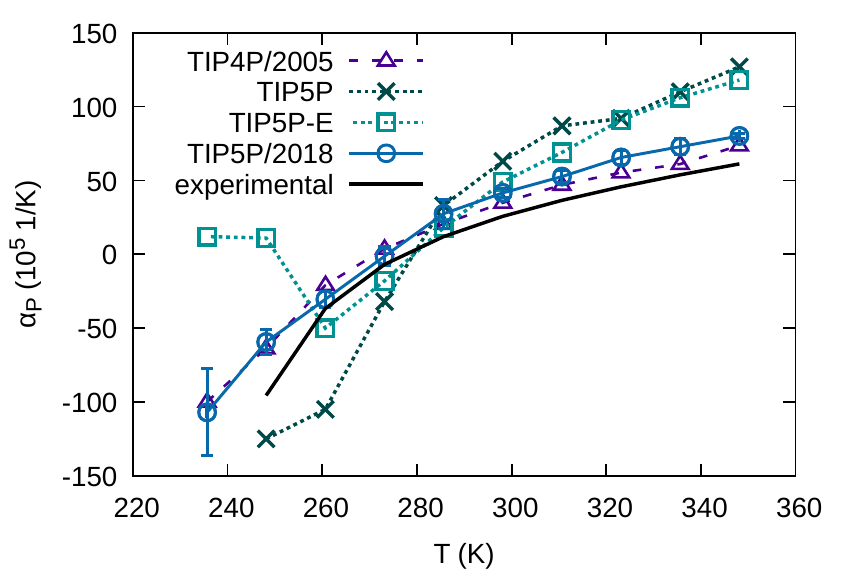}
    \caption{Coefficient of thermal expansion as a function of temperature at a pressure of \SI{1}{atm}.
	     TIP5P and TIP5P-E values are obtained from Refs. 
	     \onlinecite{mahoney_five-site_2000,rick_reoptimization_2004}, respectively, while
	     experimental data is taken from Ref. \onlinecite{kell_density_1975}.
	     Values for TIP4P/2005 and TIP5P/2018 are obtained from fluctuations of enthalpy,
	     volume, and temperature in our own simulations. 
	     Error bars for TIP5P/2018 represent standard deviations of the results from five simulations.}
    \label{fig:thermo_alpha_P}
\end{figure}

\begin{figure}
    \includegraphics[width=\columnwidth]{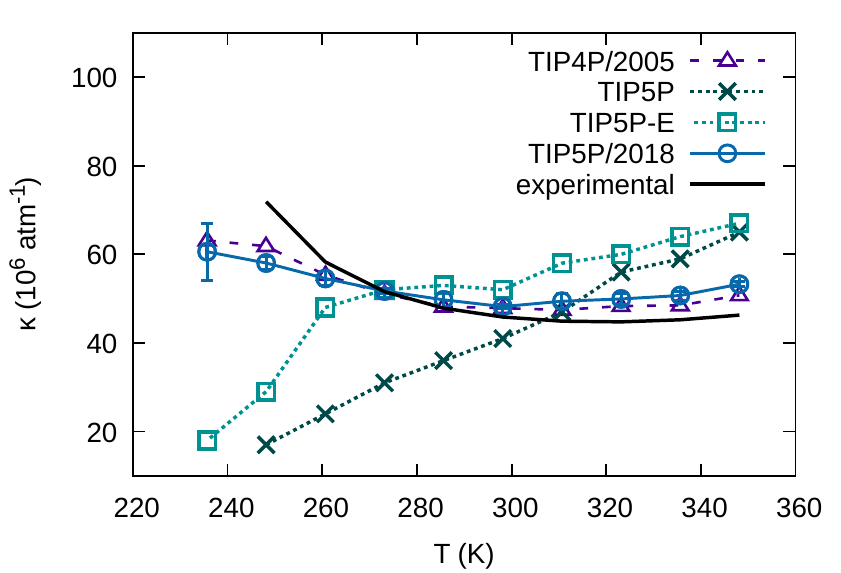}
    \caption{Isothermal compressibility as a function of temperature at a pressure of \SI{1}{atm}.
	     TIP5P and TIP5P-E values are obtained from Refs.
	     \onlinecite{mahoney_five-site_2000,rick_reoptimization_2004}, respectively, while
	     experimental data is taken from Ref. \onlinecite{kell_density_1975}.
	     Values for TIP4P/2005 and TIP5P/2018 are obtained from fluctuations of
	     volume and temperature in our own simulations. 
	     Error bars for TIP5P/2018 represent standard deviations of the results from five simulations.}
    \label{fig:thermo_K}
\end{figure}

\begin{figure}
    \includegraphics[width=\columnwidth]{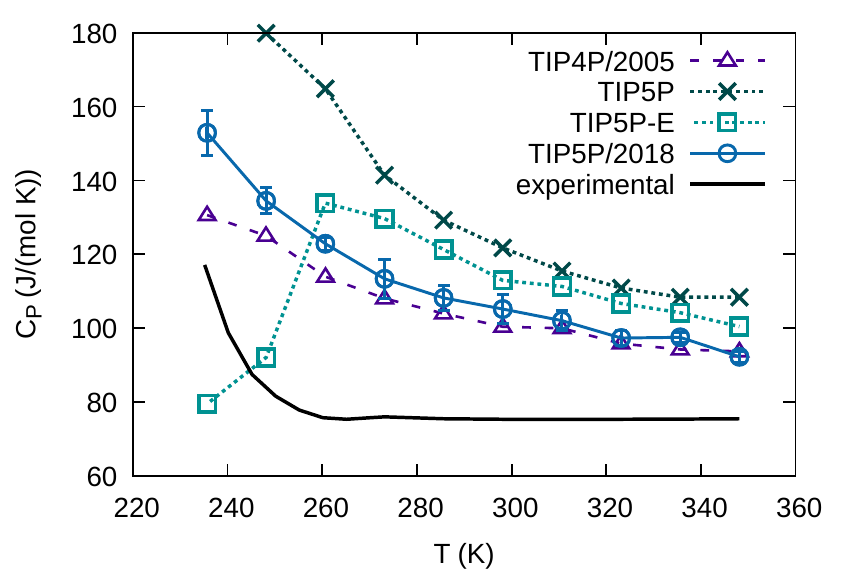}
    \caption{Isobaric heat capacity as a function of temperature at a pressure of \SI{1}{atm}.
	     TIP5P and TIP5P-E values are obtained from Refs.
	     \onlinecite{mahoney_five-site_2000,rick_reoptimization_2004}, respectively, while
	     experimental data is taken from Refs. \onlinecite{kell_notitle_1972,angell_anomalous_1973}.
	     Values for TIP4P/2005 and TIP5P/2018 are obtained from fluctuations of enthalpy
	     and temperature in our own simulations.
	     Error bars for TIP5P/2018 represent standard deviations of the results from five simulations.}
    \label{fig:thermo_Cp}
\end{figure}

The isothermal compressibility $\kappa_T$ of both TIP5P and TIP5P-E increases with temperature throughout the sampled range,
while in experiments it decreases with temperature. TIP5P/2018 reproduces the experimental trend
for temperatures down to \SI{238.15}{\kelvin}, albeit with a smaller slope (\Fig{thermo_K}).
Meanwhile, none of the five-point models are capable of reproducing the experimental
isobaric heat capacity $C_P$ (\Fig{thermo_Cp}), however TIP5P/2018 is the closest to experimental results.
Overall, the improvements TIP5P/2018 provides for the
reproduction of experimental $\alpha_P$ and $\kappa_T$ are remarkable given that
they were not used as fitting targets and instead emerge naturally from the potential.

\subsection{Rotational degrees of freedom}\label{ss:res_rot}

While the radial distribution functions and the orientational tetrahedral order distribution
provide a good description of structure, a different set of measures are required to analyze the
dynamics. To characterize translational dynamics, we have already presented measurements of diffusion.
For description of rotational dynamics of a rigid water geometry it is convenient to use rotational
autocorrelation functions
\begin{align}
 \begin{split}\label{equ:rot_auto_corr}
  C_l(t) = \bavg{P_l(\theta(t))}\,, \text{such that}
 \end{split}\\
 \begin{split}\label{equ:cos_theta_t}
  \cos(\theta(t)) = \frac{ \vec{v}(t_0+t) \cdot \vec{v}(t_0) }{ \norm{\vec{v}}^2}\,,
 \end{split}
\end{align}
where $P_l$ are Legendre polynomials of order $l$, and $\vec{v}$ is a fixed magnitude
vector the orientation of which is being studied.
Experimentally, rotational autocorrelations have been measured for molecular dipole moments
using dielectric spectroscopy\cite{bertolini_dielectric_1982,ronne_investigation_1997} ($l=1$),
and the H-H vector using nuclear magnetic resonance\cite{jonas_molecular_1976,lang_pressure_1977} ($l=2$).
In practice, these measurements are typically reported as rotational relaxation time constants,
which we calculate by fitting double exponential curves to \SI{0.25}{\nano\second} long
(\SI{2.5}{\nano\second} at temperatures below \SI{250}{\kelvin})
normalized rotational autocorrelation functions sampled every \SI{0.5}{\pico\second}.
For fitting, we use the double exponential form
\begin{equation} \label{equ:exp_exp}
a \exp \! \left[- \frac{ t}{\tau_{l}^\mathrm{type\,short}}\right] + (1-a) \exp \! \left[-\frac{t}{\tau_{l}^\mathrm{type\,long}}\right],
\end{equation}
where the superscript type corresponds to a molecular direction of either the dipole or the hydrogen-hydrogen vector (HH).
With this formulation we can extract two time scales:
one, $\tau_{l}^\mathrm{type\,long}$, corresponding to molecular reorientation due to changes in the
hydrogen bonding network and the second, $\tau_{l}^\mathrm{type\,short}$, due to
librations \cite{jimenez_femtosecond_1994,luzar_effect_1996,fecko_ultrafast_2003}.

\begin{figure}
    \includegraphics[width=\columnwidth]{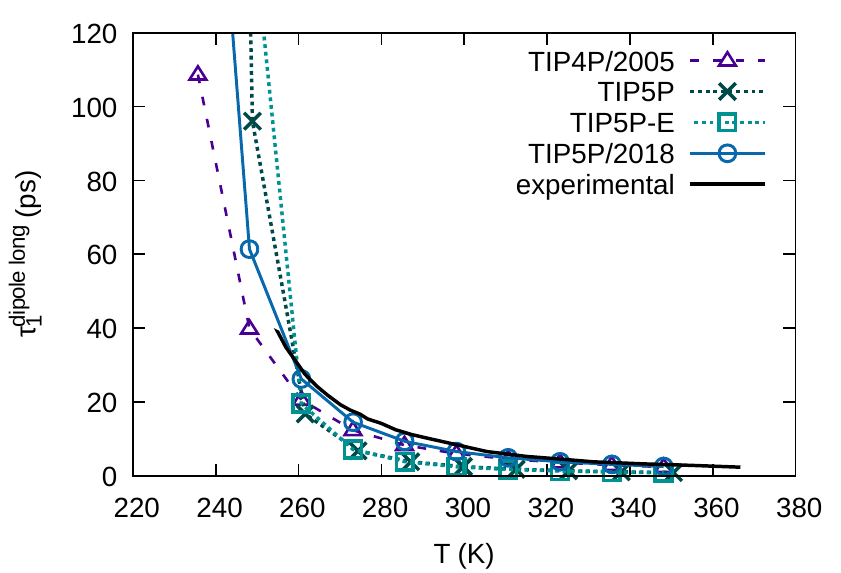}
    \caption{Temperature dependence of rotational decorrelation time $\tau_1^\mathrm{dipole\,long}$ of the dipole moment of
	     TIP5P/2018 compared to other models and experiments\cite{bertolini_dielectric_1982,ronne_investigation_1997}.
	     Values for all potentials are computed from our own simulations using
	     double exponential fits (\Equ{exp_exp}) to the autocorrelation functions.}
    \label{fig:rot_Dip}
\end{figure}

\begin{figure}
    \includegraphics[width=\columnwidth]{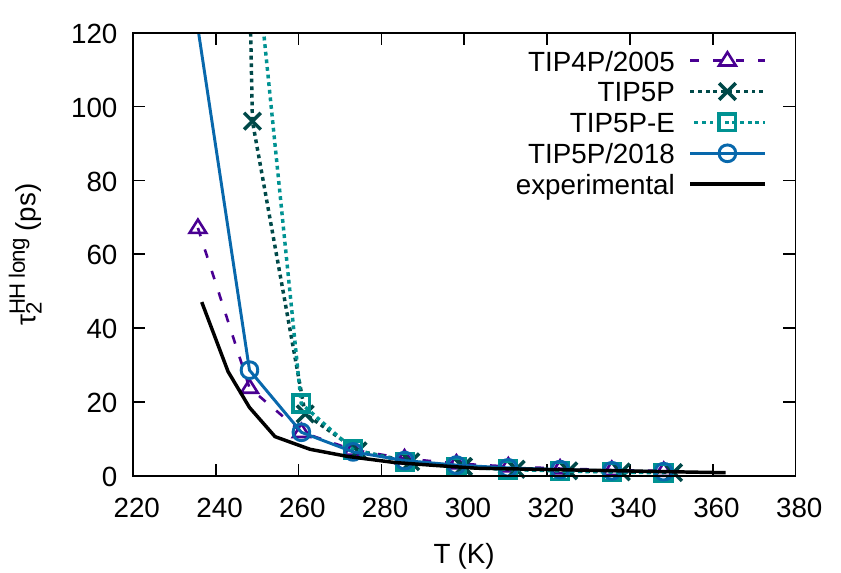}
    \caption{Temperature dependence of rotational decorrelation time $\tau_2^\mathrm{HH\,long}$ of the molecular HH vector of
	     TIP5P/2018 compared to other models and experiments\cite{jonas_molecular_1976,lang_pressure_1977}.
	     Values for all potentials are computed from our own simulations using
	     double exponential fits (\Equ{exp_exp}) to the autocorrelation functions.}
    \label{fig:rot_HH}
\end{figure}

The results show that TIP5P/2018 reproduces the experimental rotational relaxation of the molecular
dipole moment better than TIP5P and TIP5P-E (\Fig{rot_Dip}) with values of
$\tau_1^\mathrm{dipole\,long}=\SI{6.58}{ps}$, \SI{5.78}{ps}, and \SI{5.52}{ps} for
the three potentials, respectively, at \SI{298.15}{\kelvin}, \SI{1}{atm}.
A linear interpolation of experimental data\cite{ronne_investigation_1997} produces a time scale of \SI{7.69}{ps} under the same conditions.
Improvement in the rotational properties of the H-H vector are also present at lower temperatures (\Fig{rot_HH}),
but at higher temperatures TIP5P/2018 continues to slightly overestimate the experimental results with
$\tau_2^\mathrm{HH\,long}=\SI{2.90}{ps}$ at \SI{298.15}{\kelvin}, \SI{1}{atm}, against the experimental value of
approximately \SI{2.46}{ps}, as interpolated from nearby temperature points\cite{jonas_molecular_1976}.
The corresponding time scales for librations are $\tau_2^\mathrm{HH\,short}=\SI{0.24}{ps}$
and $\tau_1^\mathrm{dipole\,short}=\SI{0.50}{ps}$, however, due to the sampling time used, these values are less reliable.
As TIP5P/2018 has a dipole moment magnitude close to those of QM descriptions of water (\Tab{multipole}),
the improvement of its rotational behavior is not surprising.
Rotation of vectors perpendicular to the dipole moment, however, are also influenced by the quadrupole moment,
which TIP5P/2018 underestimates, explaining the continued deviation from experiments for the H-H vector.

\subsection{TIP5P/2018 vs. TIP4P/2005: finite size effects}\label{ss:res_4P_comp}

As TIP4P/2005 is currently the most accurate non-polarizable potential for water,
it is useful to compare TIP5P/2018 against it. The two potentials provide very similar  thermodynamic and
rotational results. The main differences in the behaviors of the two are observable in the self-diffusion
coefficients and the dielectric constants, both of which TIP4P/2005 underestimates. TIP5P/2018, on
the other hand, overestimates the dielectric constant.

\begin{figure}
    \includegraphics[width=\columnwidth]{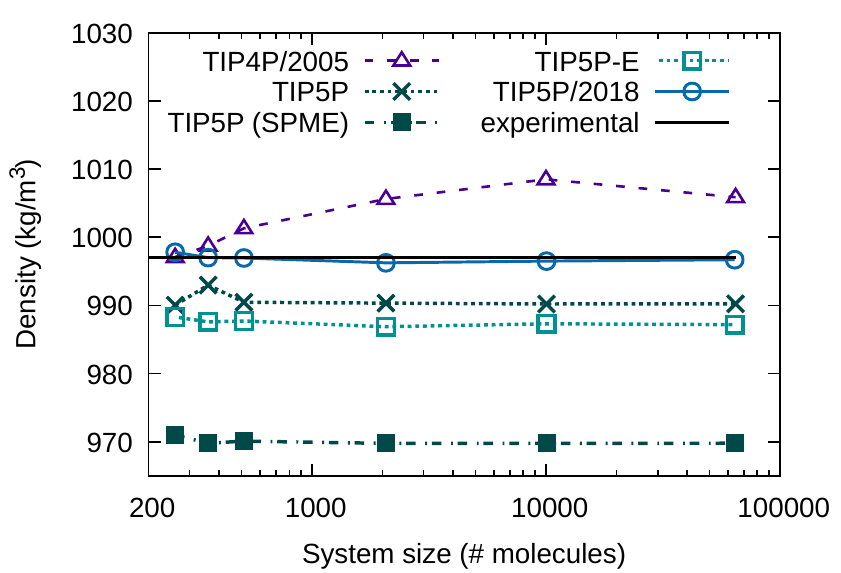}
    \caption{Density at \SI{298.15}{\kelvin} and \SI{1}{atm} as a function of system size for TIP5P/2018 and common water potentials.
    Densities are based on \SIrange{3}{20}{\nano\second} simulations with Nos\'e-Hoover thermostat\cite{nose_molecular_1984,hoover_canonical_1985-1}
    and Parrinello-Rahman barostat\cite{parrinello_crystal_1980,parrinello_polymorphic_1981-1} with \SI{1}{\pico\second} time constants.
    Experimental density\cite{Haynes2025crc} is marked with a horizontal line.
    TIP5P and TIP5P-E systems are simulated with no dispersion corrections.
    TIP5P simulated with the smooth particle mesh Ewald (SPME) method\cite{essmann_smooth_1995-1}
    is included for comparison here,
    as TIP5P combined with Ewald method\cite{ewald_berechnung_1921-1} variants is still a common sight today\cite{shen_molecular_2018,yagasaki_phase_2018}.
    The incorrect density produced by such combinations was one of the reasons for the creation of TIP5P-E\cite{rick_reoptimization_2004}.}
    \label{fig:dens_sys_size}
\end{figure}

Furthermore, having been parameterized with a 360 molecule system, TIP4P/2005 has some drift in
density as the system size increases (\Fig{dens_sys_size}). Such finite size effects are a concern
when parameterizing water potentials, as their typical use cases involve solvating biomolecules with
a large amount of water to prevent the biomolecules from interacting with their own periodic images.
To avoid such effects in TIP5P/2018, it was parameterized using 2069 molecules.
This precaution may not have been needed, however, as for both TIP5P and TIP5P-E, having been
parametrized for 512 molecule systems, our tests show little system size dependence. Overall,
TIP5P/2018 offers improved reliability over TIP4P/2005, however this comes at an increase in
computational expense of around 2.6 times.

\section {Summary and Outlook}
We have shown the viability of using \textit{ab initio} per-molecule electrostatic
potentials as a basis for charge assignment for small molecules in dense periodic systems.
We applied SQE\cite{nistor_dielectric_2009} as a more natural replacement for the charge
constraint in the RESP\cite{bayly_well-behaved_1993,cornell_application_1993} procedure.
However, further refinement of this charge assignment approach is necessary,
as not all electronic screening was taken into account.
The resulting charges proved too large to accurately capture dynamical
properties of water and uniform downscaling of the charge was required to
reproduce the experimental self-diffusion coefficient with our final potential, TIP5P/2018.

Aside from correctly capturing the maximum density of liquid water by construction,
TIP5P/2018 is also able to reproduce several emergent properties better than other
five point potentials. Improvements include thermodynamic properties, especially the drastic
improvement in the shape of the temperature dependence of isothermal compressibility,
and rotational relaxation times. It offers comparable behavior to TIP4P/2005, but
is more reliable in larger systems at the expense of an increased computational cost.

However, TIP5P/2018 still suffers in areas strongly dependent on its quadrupole moment,
as it presents with a dielectric constant significantly
higher than in experiments and possesses more preference for disordered angular
configurations than other non-polarizable potentials and \textit{ab initio} descriptions.
Therefore, further improvements to our charge assignment procedure should focus on
also reproducing the quadrupole moment of the reference \textit{ab initio}
charge distributions.

\begin{acknowledgments}

YK would like to thank Razvan Nistor for sharing his code for numerical integration of shielding interactions
and Colin Denniston for a fruitful discussion on parameterization techniques.
MK would like to acknowledge financial support from the Discovery Grants and Canada Research Chairs Program
of the Natural Sciences and Engineering Research Council (NSERC) of Canada.
This research was enabled in part by support provided by Compute Canada (www.computecanada.ca).
BB gratefully acknowledges financial support from the Innovational Research Incentives Scheme Vidi of the
Netherlands Organisation for Scientific Research (NWO) with project number 723.016.002.
BB further acknowledges the support of the NVIDIA Corporation for providing a
GTX Titan Xp GPU used for preliminary simulations.

\end{acknowledgments}


\appendix

\section{Coulomb interactions in SQE} \label{ss:appendix}

\newcommand{\Jr}[1]{{J_{#1}(r_{#1})}}
\newcommand{\JR}[1]{{J_{#1}(R_{#1})}}
\newcommand{\Ir}[2]{{I_{#1}(r_{#2})}}
\newcommand{\IR}[2]{{I_{#1}(R_{#2})}}

Coulomb interactions $U_c$ in periodic systems typically rely on Ewald
summation\cite{ewald_berechnung_1921-1}  to
include long-range contributions, $U_{\mathrm{Ewald}}$. For charge assignment, however, we
also include a shielding term $U_{\mathrm{shielding}}$ that applies
the shielded electrostatic interaction $\Jr{ij}$ only to
intramolecular atom pairs.
\begin{equation}
U_c = U_{\mathrm{ewald}} + U_{\mathrm{shielding}}
\end{equation}
\begin{equation}
U_{\mathrm{shielding}} = \frac{1}{2}\sum_{i}\sum_{j \neq i}Q_{i} Q_{j} (\Jr{ij}-1/r_{ij})
\end{equation}
The $-1/r_{ij}$ term removes the nearest image interaction introduced by Ewald summation to be replaced by
$\Jr{ij}$, an approximation fitted to reproduce the shielded electrostatic interaction between two spherically symmetrical
Slater-type orbitals\cite{slater_quantum_1960} via pairwise fitting parameters $k_{ij}$.
\begin{equation}\label{eq:Jij}
\Jr{ij}=\frac 1{4\pi \epsilon_0}\frac{2/(1+e^{-k^s_{ij}r_{ij}})-1}{r_{ij}}
\end{equation}
Atomic hardness $K_i$ of the SQE formalism can be recovered from the shielding interaction of an atom with itself,
reducing the number of free parameters in the SQE potential.
\begin{equation}
K_i = \lim\limits_{r_{ii} \to 0}\Jr{ii}
\end{equation}
For larger molecules there is no need to apply $\Jr{ij}$ to all pairs, as at larger distances it behaves as $1/r_{ij}$.


%

\end{document}